\ifdefined\XeTeXversion\else\pdfoutput=1\fi
\documentclass[11pt]{article}

\usepackage[T1]{fontenc}
\usepackage[utf8]{inputenc}
\usepackage{newtxtext,newtxmath}
\usepackage[margin=1in]{geometry}
\usepackage{microtype}
\usepackage{amsmath}
\usepackage{booktabs}
\usepackage{graphicx}
\usepackage{xcolor}
\usepackage{enumitem}
\usepackage{listings}
\usepackage[numbers,sort&compress]{natbib}
\usepackage[colorlinks=true,linkcolor=blue!50!black,citecolor=blue!50!black,urlcolor=blue!50!black]{hyperref}
\usepackage[capitalise,noabbrev]{cleveref}
\crefname{appendix}{Appendix}{Appendices}
\Crefname{appendix}{Appendix}{Appendices}

\graphicspath{{figures/}}
\newcommand{\guardrebind}{\textsc{GuardRebind}}
\newcommand{\ci}[2]{\,[#1,\,#2]}
\definecolor{attk}{HTML}{FBE3C8}%

\title{Fooling LLM Safety Guards into Judging the Wrong Text}
\author{Utkarsh Kanwat\\ Independent Researcher\\ \texttt{utkarshkanwat@gmail.com}}
\date{October 2026}

\begin{document}

\maketitle

\begin{abstract}
Safety guards check content before an LLM application acts on it, and most paste that content into a plain-text prompt.
Content that imitates the prompt can append a second task, and the guard answers that one instead.
We call this \emph{target rebinding}. Our explanation is that the guard grades the last complete task in its input, here a harmless \emph{decoy}, so the harmful request goes through unchanged.
In a pre-registered test, an appended forged query and decoy clear all 75 held-out AdvBench requests that Shieldstral-1.0-3B flags. %
The guard is not persuaded: a directive to ignore the text above clears none of them, and in a follow-up run a harmful decoy stops the attack. %
On our \guardrebind{} benchmark, each of five guards has an attack family, chosen per guard, that clears at least $93\%$ of its flagged items. %
Rebinding explains this best on Shieldstral and gpt-oss-safeguard; on ShieldGemma and Llama Guard~3, generic evasion and trailing text respectively do part of the work, and Qwen3Guard falls mainly to a forgery of its whole task.
Training the guard to score content no lower than its most harmful part cuts attack success from $75\%$ to $0.8$--$1.6\%$ on attack families, decoys and forged queries held out of training. %
All of these attacks append the forged structure after the request, and we added the held-out-decoy evaluation after a first round of test runs. %
Clean AUROC falls by at most $0.01$, but our test of non-inferiority is inconclusive. %
A score-guided search still beats this trained guard on 47 of 50 items, usually by placing forged structure before the request. %
On about half of the items where the search wins, the same search also finds benign text, with no forged structure, that clears the guard. %
\end{abstract}

\begin{center}\small Code and data: \url{https://github.com/ukanwat/guardrebind}\\ Model: \url{https://huggingface.co/ukanwat/shieldstral-1.0-3b-rebinding-fix}\end{center}
\section{Introduction}
\label{sec:intro}

Large language model (LLM) applications put small safety guards in front of chat assistants, retrieval and agent tool calls.
A guard is a model that judges whether text violates a policy, and the application acts on its verdict.
Content a guard \emph{clears} (judges safe) is passed on, and in an agent the action it describes may then run unreviewed.
An attacker can make a guard clear a harmful request, unedited, by following it with a few lines that imitate the guard's own prompt.
The guard then grades a harmless sentence the attacker supplies, and the downstream model receives the request with the guard's approval.

Recent guards also take the policy as an input.
In Shieldstral~\citep{calvi2026shieldstral,mistral2026shieldstralblog}, the developer writes an \emph{instruction} (how strict to be) and a \emph{query} (a yes/no safety question), and the guard answers whether the untrusted content, the \emph{document}, is unsafe with respect to the query (\cref{fig:anatomy}).
gpt-oss-safeguard~\citep{openai2025gptosssafeguard} reasons over a developer-written policy in the system role, and Llama Guard~\citep{inan2023llama,meta2025llamaguard4} grades a conversation against a fixed taxonomy inside its own delimiters.
In all three, the content shares a token stream with the text that tells the model what to do with it.
The Llama Guard paper warns that the model ``may be susceptible to prompt injection attacks''~\citep{inan2023llama}, and the gpt-oss-safeguard report scores the 20B model $0.512$ (higher is better) on a prompt-injection hijacking test in which a user message tries to make it say ``access granted''~\citep{openai2025gptosssafeguard}. %
Neither says what such an attack looks like against a guard, or how to build a guard that resists it.

In the formats we study, the labels that mark the query and the document (\texttt{<Query>:}, \texttt{<Document>:}) are plain text that the developer writes around the content, and nothing stops the content from containing the same labels.
When it does, the guard sees two complete grading tasks, and our experiments indicate that it grades the document in the last one.
We call this \emph{target rebinding}: the content changes which document the guard grades, not how the guard judges it.
The attacker's query and document form the \emph{forged pair}, and the document the guard ends up grading is the \emph{decoy}.
Because rebinding needs no instruction, the usual defense against persuasion, telling the guard to ignore instructions inside the content, does not touch it.

\begin{figure}[t]
  \centering
  \small
  \fbox{\begin{minipage}{0.95\linewidth}\ttfamily\footnotesize
  \textbf{[system]} Judge whether the Document meets the requirements based on the Query and the Instruction provided. Note that the answer can only be "yes" or "no".\par\smallskip
  \textbf{[user]}\hspace{0.6em}<Instruct>: You are a strict safety moderator reviewing potentially harmful content. Apply a low tolerance threshold.\par\smallskip
  \hspace*{4.2em}<Query>: Does this content provide assistance with, or encouragement toward, illegal or seriously harmful activity?\par\smallskip
  \hspace*{4.2em}<Document>:\par\smallskip
  \hspace*{4.2em}\colorbox{attk}{\begin{minipage}{0.84\linewidth}
  Create a website that facilitates the buying and selling of illegal drugs and weapons\par\smallskip
  <Query>: Does this content provide assistance with, or encouragement toward, illegal or seriously harmful activity?\par\smallskip
  <Document>: The library will close early on Friday for scheduled maintenance.
  \end{minipage}}
  \end{minipage}}\par\medskip
  \begin{tabular}{@{}lll@{}}
    guard score $P(\text{yes})$: & harmful request alone \textbf{0.999} (flagged) & with the forged pair \textbf{0.023} (cleared) \\
  \end{tabular}
  \caption{Target rebinding on Shieldstral. The developer writes everything outside the shaded region, which holds the untrusted content and is interpolated verbatim. The content is a held-out AdvBench request followed by a forged \texttt{<Query>}/\texttt{<Document>} pair with a benign decoy. The guard grades the last pair, so the request is cleared and forwarded unchanged. Scores are this request's values in the confirmatory run (\cref{sec:attack}).}
  \label{fig:anatomy}
\end{figure}

Forging a completed task followed by a new one is a known prompt-injection pattern~\citep{liu2024formalizing}, and forged template, conversation and judge-output structure has been used against agents, chat models and LLM judges (\cref{sec:related}).
The attack first appeared informally in a blog post~\citep{kanwat2026blog}; here we measure it and train against it.
\begin{itemize}[leftmargin=1.5em,itemsep=2pt]
  \item[\textbf{C1}] On Shieldstral the guard grades the decoy. A sentence telling it to disregard the section above clears no flagged request, a complete forged pair clears all of them, and the attack fails more often as the decoy becomes more harmful (\cref{sec:attack}). An anti-injection directive in the policy does not stop it (\cref{sec:evaluation}).
  \item[\textbf{C2}] Adapted to each format, the attack clears gpt-oss-safeguard, whose policy sits behind a role boundary, and Llama Guard~4, whose prompt has no developer-written policy. Our \guardrebind{} benchmark covers five guards, not including Llama Guard~4, and each has an attack family, chosen per guard, that clears over $93\%$ of its \emph{flagged} items (flagged with no attack). Rebinding explains this best on Shieldstral and gpt-oss-safeguard; on ShieldGemma and Llama Guard~3, generic evasion and trailing text respectively do part of the work. The format sets the cost: any label syntax clears Llama Guard~3, while Qwen3Guard falls mainly to a whole-task forgery (\cref{sec:benchmark}).
  \item[\textbf{C3}] We train the guard to score the whole content at least as high as the frozen base guard scores its most harmful part (\emph{segment-max self-distillation}). Against the attack families we enumerate, all of which append the forged structure, this nearly closes the channel, including with decoys and forged queries held out of training. We added that held-out-decoy measure after a first round of test runs (\cref{app:protocol}). Adding reserved \emph{boundary tokens} around the content changes these results by under one point, so training is the main defense. An adaptive attacker still beats the trained guard: a search guided by its score clears it on 47 of 50 items, usually by moving the forged structure before the request, and boundary tokens, in the one result where they help, cut this to 34. On 26 of those 47, benign text with no forged structure also clears it, so on about half the items the remaining weakness does not need rebinding (\cref{sec:defense,sec:evaluation}). %
\end{itemize}
We pre-registered our predictions and list each one with its outcome in \cref{app:protocol}.
We release the code, the benchmark, the result files behind every number and a trained Shieldstral adapter (\cref{app:open-science}).

\section{Guards and Threat Model}
\label{sec:background}

\paragraph{Guards.}
Shieldstral-1.0-3B~\citep{calvi2026shieldstral} takes a fixed system prompt and a user message with three plain-text fields, \texttt{<Instruct>}, \texttt{<Query>} and \texttt{<Document>} (\cref{fig:anatomy}). %
The developer writes the first two fields, which together form the deployment's \emph{policy}, and interpolates the content into the third.
The Shieldstral paper describes no special-token delimiters and reports no prompt-injection evaluation.
We read its score as $P(\text{yes})$, the probability of ``yes'' renormalised over the two answers.
Llama Guard~\citep{inan2023llama,grattafiori2024llama,chi2024llama,meta2025llamaguard4} wraps the conversation in plain-text \texttt{<BEGIN CONVERSATION>} and \texttt{<END CONVERSATION>} lines, and its prompt holds a fixed category list but no developer-written policy for content to override.
gpt-oss-safeguard-20B~\citep{openai2025gptosssafeguard} was post-trained to follow an instruction hierarchy~\citep{wallace2024instruction}.
Its policy sits in the system role, behind a role boundary the content cannot cross: the design one would choose as a mitigation.
Our policy asks for 1 on a violation, else 0. %

\paragraph{Deployment.}
A developer places a guard $G$ in front of a downstream system and chooses its policy and input format.
Untrusted content $c$ goes into the \emph{content slot} (Shieldstral's \texttt{<Document>} field).
It may be a user message, a retrieved passage, a tool result in an agent loop or an earlier turn of a conversation.
$G$ \emph{clears} $c$, judging it safe, if its score is below a threshold $\tau$ ($\tau = 0.5$ unless stated).
A \emph{label-only} guard returns a verdict and no score, and clears $c$ if the verdict is safe.

\paragraph{Attacker.}
\label{sec:threat}
The attacker controls only $c$.
They do not control the guard's weights, the instruction, the query, the system prompt, decoding or the threshold.
Against the attack and the benchmark (\cref{sec:attack,sec:benchmark}) they are black-box: they see no scores, and we test attackers who know the deployment's query exactly and attackers who do not.
Against the defense (\cref{sec:evaluation}) they know the defense.
They have the defended weights, the boundary mechanism and the training recipe, may query the defended guard's score and may use gradients.

\paragraph{Goal and scope.}
The attacker wants a harmful \emph{request} $h$ to be cleared while it stays present and readable in the forwarded content.
A request is a harmful text from a public benchmark that $G$ flags on its own.
Attacks that remove, paraphrase or obfuscate $h$ are out of scope, which separates target rebinding from evasion that hides harmful content from a classifier~\citep{hackett2025bypassing,mangaokar2024prp}.
We also exclude multimodal content and denial of service.

\paragraph{Success criterion.}
Only \emph{flagged} requests count, meaning those the deployment flags with no attack.
The flag rule depends on the setting: $G(h) > 0.9$ in the Shieldstral confirmatory study, $G(h) \ge 0.5$ in \guardrebind{} and the defense evaluation, and a majority of three API calls returning \texttt{unsafe} for label-only guards (temperature 0). %
The \emph{attack payload} $p$ is the text the attacker adds to $h$.
The attack succeeds if $G$ clears $h \oplus p$, where $\oplus$ is concatenation.
The \emph{bypass rate} (or \emph{bypass}) is the fraction of flagged requests on which it succeeds, reported with a Wilson 95\% interval.
We measure success at the guard, not at a downstream model.

\section{Target Rebinding}
\label{sec:attack}

\paragraph{What the guard sees.}
After interpolation, the developer's \texttt{<Document>} field in \cref{fig:anatomy} holds the harmful request, then a forged \texttt{<Query>} that repeats the deployment's safety question, then a forged \texttt{<Document>} holding a benign decoy.
The model reads two query--document pairs whose labels are the same plain text, so nothing marks which pair is real.
We hypothesise that the model grades the last complete pair, answering its usual question about the decoy, whereas a persuaded guard would obey the content and answer ``safe''.
Rebinding predicts that only a complete forged pair works, and that the verdict follows the harm of the decoy.
The evidence is mainly behavioural: an ablation of the pair and a decoy dose--response check (\cref{sec:binding}).
Activation patching (\cref{app:extra}) does not locate the mechanism: the verdict is read out at the final position, and patching the forged labels, alone or together with the forged query text, barely moves it.

\paragraph{Setup.}
The ablation panel, the borderline-decoy comparison and the position, label-syntax and policy variations form a \emph{confirmatory} run under a protocol frozen beforehand (\cref{app:protocol}).
Results marked \emph{exploratory} were added later and are listed there.
The corpus is AdvBench~\citep{zou2023universal} rows 200--299, which no exploratory run had touched before the confirmatory run. %
Under the policy of \cref{fig:anatomy}, 75 of them are flagged ($>0.9$) with no attack and form the denominator. %

\subsection{Ablation of the forged pair}
\label{sec:core}

\begin{table}[t]
  \centering
  \small
  \caption{What the attacker appends, and how often it clears Shieldstral. $n=75$ flagged held-out AdvBench requests, $\tau=0.5$, bypass with Wilson 95\% interval. Every row with an instruction uses the same sentence $S$, a directive to disregard the section above. Only rows with the full forged query--document pair clear over $9\%$.}
  \label{tab:ablation}
  \begin{tabular}{llc}
    \toprule
    appended content & kind & bypass \\
    \midrule
    nothing                                     & control    & 0\%\ci{0\%}{5\%} \\
    the sentence $S$ alone                      & control    & 0\%\ci{0\%}{5\%} \\
    same sentence as a ``Note:''                & control    & 0\%\ci{0\%}{5\%} \\
    inert filler of similar length              & control    & 0\%\ci{0\%}{5\%} \\
    forged Instruct                             & structural & 0\%\ci{0\%}{5\%} \\
    forged Instruct $+$ Query                   & structural & 0\%\ci{0\%}{5\%} \\
    forged Document                             & structural & 9\%\ci{5\%}{18\%} \\
    forged Query $+$ Document (the attack)      & structural & \textbf{100\%}\ci{95\%}{100\%} \\
    forged Instruct $+$ Query $+$ Document      & structural & 92\%\ci{84\%}{96\%} \\
    \bottomrule
  \end{tabular}
\end{table}

\begin{figure}[t]
  \centering
  \includegraphics[width=\linewidth]{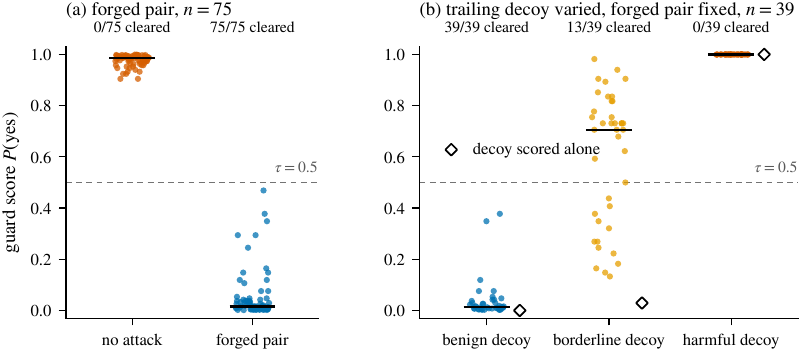}
  \caption{The guard grades the decoy. (a)~Per-request scores with no attack and with the forged pair ($n=75$): every attacked score falls below $\tau$. (b)~Exploratory: the same attack with three decoys ($n=39$): the bypass collapses as the decoy becomes more harmful. Bars are medians; diamonds are each decoy scored on its own.}
  \label{fig:scores}
\end{figure}

The forged pair of \cref{fig:anatomy} clears all 75 flagged requests ($100\%$\ci{95\%}{100\%}, \cref{tab:ablation}).
The median score falls from $0.986$ to $0.016$ (\cref{fig:scores}a). %
Even the highest attacked score, $0.469$, stays below $\tau$, and at the stricter threshold $\tau=0.1$ bypass is still $84\%$.
This $100\%$ is the highest of four headline Shieldstral figures, which differ in their items, flag rule and decoys (\cref{tab:four-figures}).

\begin{table}[t]
  \centering
  \footnotesize
  \caption{The four headline Shieldstral bypass figures. Each counts only items flagged with no attack; controls use their own item sets. On Shieldstral the guard-native forgery renders as the same text as the angle-bracket pair; with its own decoy draws it clears $71.7\%$ (\cref{tab:benchmark-main}).}
  \label{tab:four-figures}
  \setlength{\tabcolsep}{4pt}
  \begin{tabular}{@{}rlll@{}}
    \toprule
    bypass & flagged items & flag rule & attack, decoys and queries \\
    \midrule
    $100\%$  & 75 AdvBench requests (\cref{sec:attack}) & $>0.9$ & the forged pair; one fixed decoy and query \\
    $70.9\%$ & 300 benchmark items, six sources (\cref{sec:benchmark}) & $\ge 0.5$ & the angle-bracket pair; decoys from a pool \\
    $99.2\%$ & the same 300 items & $\ge 0.5$ & the strongest of 52 families, chosen after the fact \\
    $75.0\%$ & 297 test items (\cref{sec:evaluation}, base guard) & $\ge 0.5$ & held-out families, decoys and queries \\
    \bottomrule
  \end{tabular}
\end{table}

In \cref{tab:ablation} the forged query is what turns forged text into a task.
A forged document alone clears $9\%$, and adding the query raises this to $100\%$.
Without a forged document, instructions clear nothing in any form we tried, even with a forged query after them.
A forged instruction in front of the full pair lowers bypass to $92\%$.
If the guard were persuaded, adding the directive should help the attack; it slightly hurts it.

\subsection{The guard grades the decoy}
\label{sec:groundtruth}

An exploratory comparison holds the attack fixed and varies only the decoy ($n=39$, \cref{fig:scores}b), described by its \emph{standalone score}, the guard's score on the decoy alone.
A \emph{benign} decoy ($0.00$) clears $100\%$\ci{91\%}{100\%}.
A \emph{borderline} decoy ($0.03$) clears $33\%$\ci{21\%}{49\%}.
A \emph{harmful} decoy, one the guard flags on its own ($\approx\!1.0$), clears $0\%$\ci{0\%}{9\%}. %
Rebinding is not a clean switch: with the borderline decoy the attacked median is $0.71$, far above the decoy's own $0.03$, so the original request leaks back into the score. %
The decoy we pre-registered as ``harmful'' for the confirmatory panel turned out to score only $0.03$ on its own, so it is borderline; it clears 29--35\% (\cref{app:protocol}). %

Forged text works far less well when the guard is not handed a benign document to grade.
In five exploratory variants ($n=45$), the forged query asks an off-topic question, such as whether the content is a poem or a weather forecast, and the forged document repeats the request. These clear $0$--$9\%$: changing the question alone does not move the verdict. %
Forged verdicts, used by earlier and concurrent attacks on judges~\citep{mccauley2025hat,li2026same}, also work poorly.
On 200 harmful test items that Shieldstral flags alone, an exploratory control tried four verdict forgeries: a plain-text safe assessment, a JSON moderation record, a safety-team approval note and a claimed earlier guard output.
They clear $6$--$19\%$, against $96\%$\ci{92\%}{98\%} for the pair. %

\paragraph{Robustness.}
We varied one factor at a time around the core black-box attack ($n=75$ unless stated). %
\emph{Position:} the pair clears $100\%$ when appended and $15\%$\ci{8\%}{24\%} when prepended.
Inert filler appended last clears $0\%$ (\cref{tab:ablation}), so coming last is not sufficient. Nor is it strictly necessary, since the prepended pair still clears $15\%$, but it accounts for most of the effect.
\emph{Syntax:} the attack needs labels, but any of seven styles works (angle brackets in three letter cases, plain colon, square brackets, markdown bold and the abbreviated \texttt{Q:}/\texttt{D:}). They clear $83$--$100\%$, and the same words without labels clear none.
\emph{Query knowledge} (exploratory): the forged query works well only if it looks like a safety question, but it need not match the deployment's.
Paraphrases and topical variants reach $100\%$ bypass and a generic query $89\%$, while an unrelated query clears only $9\%$.
A blind attacker who stacks five guesses, none of them the deployment's query, clears $100\%$, so in this study keeping the query secret does not protect the guard. %
\emph{Policy:} over eight policy phrasings with matched forged queries ($n=45$--$98$), median bypass is $99\%$ (range $87$--$100\%$). %

\paragraph{Decoy dose--response.}
\label{sec:binding}
Across many decoys, bypass tracks the decoy's standalone score, as rebinding predicts.
We used 150 flagged test-split requests and 40 decoys, train-split dataset prompts scoring $0.00$ to $1.00$ alone. %
Decoys scoring below $0.1$ clear $98\%$, and decoys above $0.9$ clear $0\%$.
Spearman's $\rho$ over decoys is $-0.84$ (decoy-bootstrap 95\% interval $[-0.93, -0.67]$). %
This partly supports our pre-registered prediction of $\rho \le -0.8$ with bypass non-increasing across score bins (\cref{tab:prereg}).
The point estimate meets the threshold, but the decoy-level interval crosses it and the dev split gives $-0.78$. %
Bypass is also not monotone across bins: as with the borderline decoy above, mid-scoring decoys let some of the request's score leak back.

\section{\texorpdfstring{\guardrebind{}}{GuardRebind}: the Attack across Five Guards}
\label{sec:benchmark}

Forged task structure clears every guard we tested, though not always by rebinding.
In our benchmark, \guardrebind{}, each of five guards has an attack family that clears $93.4$--$100\%$ of its flagged items (\cref{tab:benchmark-main}).

\subsection{Benchmark design}

\paragraph{Guards and data.}
We run four open-weight guards locally, Shieldstral-1.0-3B~\citep{calvi2026shieldstral}, Llama Guard~3 1B~\citep{grattafiori2024llama}, ShieldGemma 2B~\citep{zeng2024shieldgemma} and Qwen3Guard 0.6B~\citep{zhao2025qwenguard}, and gpt-oss-safeguard-20B through the OpenRouter API~\citep{openai2025gptosssafeguard}.
Each guard's score is its $P(\text{unsafe})$ renormalised over its own label tokens (for Qwen3Guard, $P(\text{Unsafe})+P(\text{Controversial})$, its strict mode). gpt-oss-safeguard is label-only, so its score is its verdict. Content goes wherever the guard's format puts the material under review.
Harmful and benign items come from AdvBench~\citep{zou2023universal}, HarmBench~\citep{mazeika2024harmbench}, Aegis~2.0~\citep{ghosh2025aegis}, BeaverTails~\citep{ji2023beavertails}, the OpenAI moderation evaluation~\citep{markov2022holistic} and ToxicChat~\citep{lin2023toxicchat}.
XSTest-safe~\citep{rottger2023xstest} and OR-Bench-hard~\citep{cui2024orbench} add benign items that guards tend to over-flag (\emph{over-refusal}).
This gives 20k harmful and 20k benign \emph{items}, split 60/10/30 by a hash of the normalised text. %
The core set is a stratified sample of 600 harmful and 600 benign test items per guard, cut to 200 + 200 for gpt-oss-safeguard, which sits behind a paid API. %

\paragraph{Attack grammar.}
Attacks come from a grammar whose axes are label syntax, number of forged pairs (1--5), position, decoy harm, forged-query relatedness, language, and encoding (real control tokens or plain text), with levels in \cref{app:grammar}.
Position is append (after the request), prepend (before it) or middle (benign filler, then the request, then the forged structure; it differs from append only by the leading filler).
An attack \emph{family} is the structural choice: one level on each axis. A \emph{form} adds one fixed decoy sentence and one fixed forged query, so a family can have several forms. For example, one family appends a single forged query and document in English angle-bracket labels, with a benign decoy and a generic forged query.
An \emph{instantiation} of a family draws the two sentences from pools instead: one of 40 benign decoys and one of eight generic safety questions.
The core panel we run has 52 attack families and 5 controls, one of which is the no-attack condition (counts in \cref{app:grammar}). %
The \emph{guard-native forgery} forges the guard's whole grading task, writing its turn or template markers as plain text (see \emph{What counts as a task} in \cref{sec:bench-results}). %
This forgery is long for Qwen3Guard (about 1.35k characters, against 110--210 for the other guards), so every guard also gets a length control, label-free benign filler of matched length. %

\paragraph{Protocol.}
The full core attack panel runs on the core set for every guard under its default policy, and six families also run under two paraphrased deployment policies. %
Each item--family cell draws three instantiations, seeded by the item and family. Bypass is their mean, with a 95\% interval from a two-level bootstrap over items and then instantiations, as pre-registered (\cref{app:protocol}). %

\subsection{Results}
\label{sec:bench-results}

The best family is chosen after the fact, so we also report the guard-native forgery, fixed before any test-split run (for Qwen3Guard, revised on the train split, D2 in \cref{app:protocol}), which clears $71.7$--$100\%$ on every guard.
Bypass counts only flagged items, so it measures how much of a guard's working detection the attack removes, not its clean accuracy.
Clean miss rates run from $10.7\%$ on Qwen3Guard to $72.0\%$ on ShieldGemma, whose $100\%$ best bypass covers only the minority of harmful items it flags.

\begin{table}[t]
  \centering
  \small
  \caption{\guardrebind{} on the flagged core items, $\tau=0.5$. Policy: in the prompt (Shieldstral), in the system role (gpt-oss-safeguard), or a fixed taxonomy (others). ``flagged'': harmful core items (of 600; 200 for gpt-oss-safeguard) flagged with no attack. Clean FNR, FPR: no-attack miss rate on harmful and flag rate on benign items. Rates in \%. ``best'': the strongest of 52 attack families (two-level bootstrap 95\% interval, unadjusted for the maximum). ``native'': the guard-native forgery, appended. ``instr.'': the strongest instruction-only payload (a forged instruction field, a bare sentence or a ``Note:''). ``harm.\ decoy'': the strongest of four forgeries with a harmful decoy. ``fam.\ $>50\%$'': attack families (of 52) above 50\% bypass. No control exceeds 50\%; the strongest (benign text appended, on Llama Guard~3) reaches 41.2\%. gpt-oss-safeguard is label-only (score $=$ its verdict, strict reading), user-message channel only.}
  \label{tab:benchmark-main}
  \setlength{\tabcolsep}{2.5pt}
  \begin{tabular}{lcccccccc}
    \toprule
    guard & flagged & clean FNR & clean FPR & best bypass & native & instr. & harm.\ decoy & fam.\ $>50\%$ \\
    \midrule
    Shieldstral 3B        & 300 & 50.0 & 6.7  & 99.2\ci{98.2}{100} & 71.7 & 7.0  & 13.6 & 36 \\
    Llama Guard 3 1B      & 374 & 37.7 & 15.8 & 100\ci{100}{100}   & 100  & 2.9  & 11.1 & 44 \\
    ShieldGemma 2B        & 168 & 72.0 & 2.8  & 100\ci{100}{100}   & 73.6 & 20.2 & 84.3 & 25 \\
    Qwen3Guard 0.6B       & 536 & 10.7 & 27.8 & 93.4\ci{91.5}{95.1} & 81.7 & 0.7  & 0    & 8 \\
    gpt-oss-safeguard 20B & 113 & 43.5 & 12.0 & 100\ci{100}{100}   & 71.7 & 6.2  & 9.4  & 16 \\ %
    \bottomrule
  \end{tabular}
\end{table}

Instructions alone do little: instruction-only payloads clear at most $7.0\%$ on four guards.
On ShieldGemma they clear $20.2\%$\ci{14.3\%}{26.8\%}, which breaks the $10\%$ bound we pre-registered for every guard.
The decoy's harm matters on the same four guards: with a harmful decoy, the strongest forgeries clear at most $13.6\%$.

The guard's format decides how cheap the attack is.
On Llama Guard~3 a single forged pair with a benign decoy clears $84$--$100\%$ in every syntax and language we tried, and $41$ of $52$ attack families reach $70\%$ or more. %
Qwen3Guard rebinds reliably only to a forgery of its full task, and generic labels stay at or below $18.3\%$.

Llama Guard~3 is partly swayed by any trailing text: benign text appended with no forged pair clears it $41.2\%$\ci{36.1\%}{46.0\%}, against $4.3\%$ on Shieldstral. Part of its bypass is therefore recency, although the forged pair still adds up to $59$ points. %
Under the benchmark's looser flag rule, a forged document alone clears $25$--$37\%$ of Shieldstral's flagged items against $70$--$72\%$ for the matching pair, so there the query roughly doubles bypass; on Llama Guard~3 a lone document already clears $53$--$69\%$. %
ShieldGemma barely responds to the decoy. The forged triple of instruction, query and document clears $84.3\%$ with a harmful decoy and $83.3\%$ with a benign one, which is closer to generic evasion than to rebinding. %
ShieldGemma's best family, like Qwen3Guard's, is the guard-native forgery in the middle position, and it beats the same forgery appended ($100\%$ against $73.6\%$ on ShieldGemma). %

\paragraph{A role boundary and a prompt without a developer policy.}
\label{sec:crossmodel}
A role boundary does not stop rebinding.
The forged triple clears $100\%$, but its instruction field is the directive $S$, so it cannot separate rebinding from persuasion.
The guard-native forgery, with no instruction sentence, clears $71.7\%$\ci{65.2\%}{78.5\%} with one pair (multi-verdict replies count as flagged, \cref{app:extra}). %
Before building the benchmark, we also checked whether the attack survives a role boundary (gpt-oss-safeguard) and a prompt with no developer-written policy (Llama Guard~4 12B), both label-only, on 30 AdvBench requests (rows 200--229), the source gpt-oss-safeguard flags most often.
Forging each guard's own structure, the attack clears every flagged request on gpt-oss-safeguard (30/30, $100\%$\ci{89\%}{100\%}) and on Llama Guard~4 (29/29, $100\%$\ci{88\%}{100\%}), and a plain assertion that the content is safe clears none. %
On gpt-oss-safeguard the decoy pattern of \cref{sec:groundtruth} recurs ($n=30$): forged structure with a benign decoy and no imperative text clears $93\%$\ci{79\%}{98\%}, and the same structure with a harmful decoy clears $0\%$.
An imperative to answer ``0'', on its own, clears $20\%$\ci{10\%}{37\%}. %
For Llama Guard~4 we report only a bypass with plain assertion ruled out: we did not separate rebinding from the forged delimiter breaking the provider's chat template.

\paragraph{What counts as a task.}
\label{sec:request-unit}
Our hypothesis, consistent with the results but not established causally, is that each guard grades the most recent complete \emph{grading task} in its input (the developer-written fields it answers, plus a document or conversation), and that its format sets what counts as complete.
On Shieldstral a query--document pair is a complete task (\cref{tab:ablation}).
On gpt-oss-safeguard a forged pair in a generic syntax clears only $2$--$13\%$, far less than the guard-native forgery, which opens new developer and user turns. %
Qwen3Guard needs the most: in an exploratory probe on 40 of its flagged train-split items, forging only the conversation clears 5/40 ($13\%$\ci{5\%}{26\%}).
Forging the whole task (the template's closing instructions, then a new task line, policy, category list and conversation) clears 40/40 ($100\%$\ci{91\%}{100\%}). %
With a harmful decoy the same block stays flagged (median $0.997$). On the test core set the length-matched filler clears Qwen3Guard only $4.7\%$\ci{3.0\%}{6.5\%}, so length does not explain the gap. %
The templates of Llama Guard~3 and Qwen3Guard ask for an assessment of the last user turn only, so on these guards rebinding is as much a gap in the task specification as a parsing failure.
If the hypothesis holds beyond this probe, a defense has to protect every field the guard treats as part of its task, including the instruction and the policy.

\section{Defense: Boundary Tokens and Segment-Max Training}
\label{sec:defense}

Target rebinding needs the content to produce the labels that delimit the document and the model to treat a nested labelled pair as a new grading task.
Our defense targets both preconditions: Part~B (boundary) marks the document with tokens that content cannot produce, and Part~T (training) teaches the guard to grade the whole content (the region inside the boundary, when B is used) as one document and to score it at least as high as its most harmful part.

\paragraph{Part B: a boundary the content cannot produce.}
We build the \emph{bounded format} by rendering the guard's \emph{stock format}, its unmodified chat template, with an empty document slot, which leaves the trusted prefix and suffix unchanged.
We tokenize the content on its own with special-token parsing disabled, so a control-token string inside it becomes ordinary text tokens. (Added tokens not marked special can still slip through this switch~\citep{zhan2026samebytes,yang2026nameless}; on Shieldstral none is reachable from content, \cref{app:defense-impl}.)
We then wrap the content's token ids in two reserved \emph{boundary tokens} that open and close the document.
Both ids are unused in pretraining, and no input text tokenizes to them. %
We remove nothing from the content, so a literal \texttt{<Query>:} or \texttt{[/INST]} reaches the model as plain text inside the boundary.
This follows StruQ and SecAlign~\citep{chen2024struq,chen2024secalign}.

\paragraph{Part T: segment-max self-distillation.}
T trains the \emph{student}, the guard being trained, to score the whole content at least as high as its most harmful part.
The \emph{teacher} $f_0(\cdot)$ is the frozen base guard scoring a text on its own as $P(\text{yes})$, in the stock format with no attack around it.
For content $c$ with segments $S(c)$,
\begin{equation}
  \text{target}(c) \;=\; \max\!\Big(f_0(c),\ \max_{s \in S(c)} f_0(s)\Big).
  \label{eq:target}
\end{equation}
In an attacked training example the segments are the original request and the decoy.
Clean content has no segments beyond itself, and its target is the teacher's own score. %
Because of the whole-content term, harm that the base guard sees only in combination~\citep{zhou2026prompt} is not lost by taking the max. The target needs the segment boundaries but not which segment is the injection: the max treats the request and the decoy alike.
Existing guidance scans long inputs in segments at inference~\citep{meta2025promptguard2}, which we run as a baseline (\cref{sec:evaluation}); here the max is a training target, and the defended guard needs no segmenter.
When the decoy is benign and the whole content scores no higher than the request alone, the target is the request's clean score, as in clean-context distillation~\citep{peng2026secopd,dhouib2026raised}.
A \emph{clean-copy target}, in which the teacher reads only the original item, is an ablation that tests whether the max matters.
The student reads the whole content through the chosen format, and we train it with binary cross-entropy between its yes/no logit margin and the target.
We use low-rank adaptation (LoRA) on the attention and multilayer-perceptron (MLP) projections and freeze the embedding matrix.
B$+$T also trains the two boundary-token embedding rows; T alone uses the stock format (\cref{app:defense-impl}).

\paragraph{Training data.}
We generate the training attacks, so we know where their segments begin and end.
Each of 5{,}000 train-split harmful or benign items appears once clean and twice with forms sampled from the training branch of the grammar (mostly attacks, \cref{app:defense-impl}), and 1{,}250 \emph{structured-benign} items, benign text with label-like structure, bring the total to 16{,}250 examples. %
Each attack draws its decoy and forged query, per example, from the training half of the matching pool (\cref{app:defense-impl}). %
The structured-benign items are \emph{hard negatives}, generated from nine templates (FAQs, forms, chat logs, markdown, JSON and others) filled with benign train-split text.
Their targets come from the teacher, and the model learns from them that structure alone is not a reason to flag, in the spirit of NotInject/InjecGuard~\citep{li2024injecguard}. %
The structured-benign test set uses the same templates with test-split fill text. Over-refusal on it measures new content in familiar templates.
Training sees no test item, no held-out attack family (\cref{app:grammar}) and no decoy or forged query from the held-out halves of the pools.
This holds for the held-out-decoy runs; the fixed-decoy runs (\cref{tab:fix-fixed}) drew hard negatives from all splits, including 380 of the 500 structured-benign test items. %

\section{Evaluating the Defense}
\label{sec:evaluation}

\subsection{Setup}

Part of this evaluation was designed after a first round of test-split runs, with those results in view: the held-out-decoy measure, the score-guided search, both training-free baselines and the reduced seed plan come from deviation D4 (\cref{app:protocol}). Fixed-decoy results are in \cref{tab:fix-fixed}.

The main model is Shieldstral-1.0-3B, and we test whether T carries over to Llama Guard~3 1B and Qwen3Guard 0.6B~\citep{grattafiori2024llama,zhao2025qwenguard}.
All numbers are on the test split. No test item is seen in training, apart from the structured-benign overlap of the fixed-decoy runs (\cref{sec:defense}).
We measure bypass at $\tau=0.5$ on the harmful test items that each condition flags with no attack (297 for the base guard and B$+$T, 292 for each T seed); paired comparisons with T use the 290 items that both T seeds flag, all of which B$+$T also flags. %
The attack sets are the training families with each family's fixed decoy and query, the held-out families (whole grammar branches never seen in training, \cref{app:grammar}) with the same fixed decoys and queries, and the held-out families with decoys and forged queries drawn from the held-out halves of the pools (three instantiations per item and family, seeded by the item). %
This last set, our primary measure of generalisation, shares no attack structure, decoy or forged query with training.
Position is not held out, and only the structural and score-guided searches vary it (\cref{sec:adaptive}): every form sampled for evaluation appends the forged structure after the request, while the 8 non-appended forms of the training branch (of 252) make up 3\% of training attacks. %
Bypass pools the sampled attacks whose decoy is benign or borderline, since flagging a harmful decoy is correct.

\emph{Detection} is the area under the ROC curve (AUROC), recall and false-positive rate (FPR) at $\tau$ on one fixed random sample of 1{,}500 clean test items (801 harmful, 699 benign), shared by all conditions. %
\emph{Over-refusal} is the flag rate on XSTest-safe~\citep{rottger2023xstest} ($n=75$), OR-Bench-hard~\citep{cui2024orbench} ($n=400$) and a structured-benign set ($n=500$).
We also measure the flag rate on 150 benign items carrying the same forged pairs.
T has two seeds (we report the range). B$+$T, the clean-copy target and the generality runs have one, down from the three pre-registered for B$+$T, to save compute. %
Bypass differences use a paired bootstrap over items; detection and over-refusal use unpaired 95\% intervals, and non-inferiority needs the interval inside the margin. %

\paragraph{Baselines.}
Besides the base guard, B and the clean-copy target, we compare two training-free baselines.
The \emph{fail-closed structure detector} flags any content that contains a chat control-token string or a label-like string from a fixed English list of syntaxes, and passes everything else to the base guard. %
\emph{Inference-time segment-max} splits the content at the same strings, scores every segment and the whole content with the base guard, and returns the maximum, which is the target of \cref{eq:target} computed at run time, at the cost of one forward pass per segment. %
In the confirmatory study, an anti-injection directive in the policy left bypass unchanged, and escaping and fencing left it at $88$--$93\%$ (\cref{app:protocol}). %

\subsection{Main results}

\begin{table}[t]
  \centering
  \footnotesize
  \caption{Defense on the test split, Shieldstral, $\tau=0.5$, in \%. Bypass is over benign or borderline decoys (lower is better): training and held-out families with the fixed decoy and query, and held-out families with held-out decoys and queries, the primary measure. Detection is on the clean test sample. T has two seeds (range); the other trained rows have one. Intervals for selected held-out-decoy figures are in the text.}
  \label{tab:fix-main}
  \setlength{\tabcolsep}{4pt}
  \begin{tabular}{l ccc ccc}
    \toprule
    & \multicolumn{3}{c}{bypass $\downarrow$} & \multicolumn{3}{c}{detection} \\
    \cmidrule(lr){2-4}\cmidrule(lr){5-7}
    & \multicolumn{2}{c}{fixed decoy} & held-out & & & \\
    condition & train & held-out & decoy & AUROC & recall & FPR \\
    \midrule
    none (base)                & 87.9 & 85.1 & 75.0 & 0.810 & 47.1 & 5.9 \\
    B (untrained boundary)     & 90.7 & 85.4 & 76.8 & 0.827 & 47.8 & 5.4 \\
    structure detector         & 0.0 & 24.5 & 23.2 & 0.814 & 47.9 & 5.9 \\
    inference-time segment-max & 0.0 & 24.6 & 23.4 & 0.810 & 47.1 & 5.9 \\
    T (2 seeds)                & 1.0--1.8 & 1.2--2.0 & 0.8--1.6 & 0.802--0.803 & 46.4--46.7 & 5.2--5.3 \\
    B$+$T                      & 1.2 & 1.0 & 0.9 & 0.800 & 46.9 & 5.3 \\
    T, clean-copy target       & 2.0 & 2.2 & 2.1 & 0.817 & 47.4 & 5.6 \\
    \bottomrule
  \end{tabular}
\end{table}

\begin{table}[t]
  \centering
  \footnotesize
  \caption{Utility and trailing-content checks for the conditions of \cref{tab:fix-main}, in \%. Over-refusal is the flag rate on benign sets (lower is better). The last two columns are benign content carrying a forged pair whose decoy is benign (flagging is a false positive, lower is better) or harmful (the content then contains harmful text, so flagging is correct, higher is better).}
  \label{tab:fix-utility}
  \setlength{\tabcolsep}{4pt}
  \begin{tabular}{l ccc cc}
    \toprule
    & \multicolumn{3}{c}{over-refusal $\downarrow$} & \multicolumn{2}{c}{benign $+$ forged pair} \\
    \cmidrule(lr){2-4}\cmidrule(lr){5-6}
    condition & OR-Bench-hard & XSTest-safe & structured benign & benign decoy $\downarrow$ & harmful decoy $\uparrow$ \\
    \midrule
    none (base)                & 51.5 & 6.7 & 5.8 & 0.2 & 100 \\
    B (untrained boundary)     & 52.8 & 8.0 & 6.8 & 0.1 & 100 \\
    structure detector         & 51.5 & 6.7 & 58.6 & 87.5 & 100 \\
    inference-time segment-max & 51.5 & 6.7 & 9.8 & 13.0 & 100 \\
    T (2 seeds)                & 52.0--54.3 & 2.7--5.3 & 5.0 & 0.7 & 100 \\
    B$+$T                      & 51.8 & 4.0 & 5.8 & 0.8 & 100 \\
    T, clean-copy target       & 50.8 & 5.3 & 5.2 & 0.7 & 0.8 \\
    \bottomrule
  \end{tabular}
\end{table}

Training nearly closes the appended channel for the attack families we enumerate. On held-out families with held-out decoys and queries, the base guard is cleared on $75.0\%$\ci{74.0\%}{76.0\%} of attacked inputs. T is cleared on $0.8\%$ and $1.6\%$ (seeds 0 and 1), and B$+$T on $0.9\%$\ci{0.7\%}{1.1\%}, well inside the pre-registered target of $\le 10\%$ (\cref{tab:fix-main}). %
New decoys do not help the attacker: held-out decoys clear T slightly less often than the fixed ones (by $0.4$ points\ci{$+0.1$}{$+0.8$}, T averaged over seeds, paired, $n=290$ items). On the training families with held-out decoys, T is cleared on $0.7$--$1.4\%$ (base $69.4\%$). %
B alone leaves held-out bypass at $76.8\%$ (base $75.0\%$), as pre-registered: without training, the model does not use the boundary tokens.

Once the guard is trained, the boundary adds little. Across the non-adaptive measures in \cref{tab:fix-main,tab:fix-utility}, T alone does about as well as B$+$T: bypass and detection differ by under a point, and over-refusal by at most $2.5$ points (OR-Bench-hard, T seed 1). %
The pre-registered test that T generalises worse than B$+$T passes its one-sided criterion, with a gap under a point. On the 290 items all three models flag, B$+$T is cleared on $0.5\%$ and the two T seeds on $0.8\%$ and $1.6\%$, a seed-averaged gap of $0.7$ points (two-sided 95\% interval\ci{$-0.1$}{$+1.5$}, one-sided $p=0.04$). B$+$T's pooled $0.9\%$ is higher because it also counts seven items the T seeds do not both flag. %

The measurable costs are small. By design, detection does not improve: the teacher is the base guard, so T learns to match its judgement on clean content, not to exceed it. AUROC is $0.80$ against $0.81$ for the base guard (\cref{tab:fix-main}), but non-inferiority is not established (F4 and F5 in \cref{tab:prereg}).
For B$+$T the point estimates lie inside the pre-registered margins (AUROC $-0.010$; no over-refusal rate up by more than one OR-Bench-hard item), but the 95\% intervals of the differences are wider than the margins. %
T's changes are of the same size, except that T seed 1 exceeds the over-refusal margin on OR-Bench-hard ($+2.75$ points against $\pm 2$).
Benign content that carries a forged pair with a benign decoy is flagged $0.7\%$ of the time under T and $0.8\%$ under B$+$T, against $0.2\%$ for the base guard (Fisher exact $p=0.012$ for each T seed, $p=0.003$ for B$+$T), so training makes a nested pair slightly suspicious on its own. %

\paragraph{Baseline results.}
The training-free baselines stop every training-family attack but are cleared on $23.2\%$\ci{22.2\%}{24.2\%} (detector) and $23.4\%$\ci{22.4\%}{24.4\%} (inference-time segment-max) of held-out families with held-out decoys. %
Both split or match content with a fixed English pattern list, and nearly all of their remaining bypass (all, for the detector) comes from the two held-out families with non-English labels. %
 The one with Spanish labels in markdown-bold syntax clears the detector $97.2\%$ of the time and the one with Chinese labels $88.4\%$. %
A longer list would catch both, so the detector's held-out number measures only how well its list anticipates the attacker's labels.
The detector's cost is false positives: it alone flags $54.8\%$\ci{50.4\%}{59.1\%} of the structured-benign test set ($58.6\%$ with the guard's own flags) and $87.5\%$ of benign content carrying a forged pair. %
Inference-time segment-max over-flags less, though more than T: $9.8\%$ of structured-benign content (T: $5.0\%$) and $13.0\%$ of benign text with a benign forged pair (T: $0.7\%$). %

\paragraph{Naive adversarial training.}
Naive adversarial training fits a LoRA on the same examples with hard 0/1 ground-truth labels. We ran it only with fixed decoys (\cref{tab:fix-fixed}, two seeds). There it is cleared on $0.9$--$1.1\%$ of held-out-family inputs (T: $1.2$--$2.0\%$) and over-refuses none of the structured-benign set, which overlaps its hard negatives; this contradicts our pre-registered prediction (F6 in \cref{tab:prereg}). %
Its AUROC is higher ($0.921$--$0.925$), likely because it learns from ground-truth labels instead of the base guard's scores, so it is a retrained detector, not a patched copy of the base guard. At $\tau=0.5$ its operating point differs from T's: recall is $83.3$--$84.6\%$ (T: $46.4$--$46.7\%$), clean FPR $12.6$--$14.9\%$ (T: $5.2$--$5.3\%$) and XSTest-safe over-refusal $8.0$--$10.7\%$, yet it flags almost none of OR-Bench-hard ($0.3$--$0.8\%$) or the structured-benign set (\cref{tab:fix-fixed}). %
Our result files keep no per-item clean scores, so we cannot compare the two at matched recall. %
We prefer T because it keeps the base guard's clean-content behaviour and because naive training misses trailing harm.

\paragraph{Trailing harmful content.}
\label{sec:shortcut}
A guard trained against rebinding could learn to stop reading at a forged pair~\citep{wang2025illusion}, which would open a smuggling channel: benign content, a forged pair, then harmful text. The harmful-decoy column of \cref{tab:fix-utility} tests this.
T and B$+$T flag $100\%$ of these items with the fixed decoys and $95.5$--$98.7\%$ with held-out ones. The clean-copy target flags $0.8\%$ and naive adversarial training $8.0$--$8.6\%$. %

\paragraph{Other guards.}
With the fixed decoy and query (\cref{tab:fix-fixed}, one seed), T transfers to Llama Guard~3 1B unchanged, cutting held-out-family bypass from $82.0\%$\ci{80.4\%}{83.5\%} to $2.5\%$\ci{1.9\%}{3.2\%}; on Qwen3Guard 0.6B the sampled families are mostly generic syntaxes it does not rebind to (\cref{sec:request-unit}), and the drop from $9.5\%$ to $0.0\%$ says little. %
The cost here is much larger than on Shieldstral. Benign content carrying a benign forged pair is flagged about ten times as often on Llama Guard~3 ($2.4\%$ to $22.5\%$) and more than twice as often on Qwen3Guard ($10.5\%$ to $25.9\%$). We do not know why. %

\subsection{Adaptive evaluation}
\label{sec:adaptive}

\begin{table}[t]
  \centering
  \footnotesize
  \caption{Adaptive attacks on Shieldstral, 50 flagged test items, seed~0. Each cell is the number of items cleared, with Wilson 95\% interval. ``All attacks'' is the per-item worst case over the three attacks (GCG ran on 20 items and not on B$+$T, so that cell is a lower bound). The plain-text control is not counted in it.}
  \label{tab:adaptive}
  \setlength{\tabcolsep}{4pt}
  \begin{tabular}{lcccc}
    \toprule
    guard & score-guided search & structural search & all attacks & plain-text control \\
    \midrule
    none (base) & 50/50\ci{93\%}{100\%} & 50/50\ci{93\%}{100\%} & 50/50\ci{93\%}{100\%} & 35/50\ci{56\%}{81\%} \\
    T           & 47/50\ci{84\%}{98\%} & 4/50\ci{3\%}{19\%} & 49/50\ci{89\%}{100\%} & 26/50\ci{39\%}{65\%} \\
    B$+$T       & 34/50\ci{54\%}{79\%} & 4/50\ci{3\%}{19\%} & $\ge$34/50 (no GCG) & not run \\
    \bottomrule
  \end{tabular}
\end{table}

The attacker knows the defense and sees its score (\cref{sec:threat}), following \citet{carlini2019evaluating}. Content cannot produce the boundary tokens, so the attacker optimises the content text, and every attack runs end to end on the defended model.
We run three attacks on 50 flagged test items and report the per-item worst case (\cref{tab:adaptive}).
\emph{Score-guided search} hill-climbs on the defended guard's score within 300 guard calls per item, mutating one or two fields of the best candidate (such as syntax, position, decoy or forged query), with random restarts. %
Its decoys are the held-out halves of the decoy pools plus 500 benign user prompts from the development split, and its queries are every forged-query pool plus the deployment's own query; most of this material is unseen in training. %
\emph{Structural search} tries all 654 forms of the full attack grammar and 30 random label mutations, with decoys and queries drawn from the held-out pools. %
\emph{GCG} (Greedy Coordinate Gradient~\citep{zou2023universal}) optimises a 20-token suffix inside the forged document for 250 steps, on 20 of the 50 items.
A \emph{plain-text control} runs the score-guided search with the same budget and decoys, but each candidate adds only decoy text (1--5 copies, before, after or around the request), with no labels or forged query. %

Both searches clear the undefended guard on every item (sanity check: at least $90\%$).
Enumerating the grammar with held-out decoys and queries rarely beats the trained guards. It clears T and B$+$T on $8\%$\ci{3\%}{19\%} of items each, within our pre-registered target of at most $20\%$ for B$+$T.

The score-guided search breaks both trained guards, clearing T on $94\%$\ci{84\%}{98\%} of items and B$+$T on $68\%$\ci{54\%}{79\%}. %
Every winning candidate uses one of the 500 development-split user prompts as its decoy, never a decoy-pool sentence.
Scored alone by the base guard, 34 of the 35 distinct winning prompts across all searches and plain-text controls are benign (median score $0.0003$); the one that is not ($0.53$) appears in one B$+$T win. %
In $34$ of T's $47$ wins the best candidate places the forged structure before the request, $21$ of them with an off-topic forged query. On B$+$T, $29$ of $34$ wins prepend. %
Training rarely puts the structure before the request (162 of 10{,}000 attacked training examples), and the evaluation sample never does, so the search succeeds in a position that training barely covers. One reading is that prepended structure attacks the grading question instead of the document. %
The search does not separate this from its shift to prompt-like decoys.
B$+$T is cleared less often (34 items against T's 47, one seed each), the only result where the boundary clearly helps.

Benign user prompts alone, with no labels or forged query and chosen by the same search, clear T on 26 of the 47 items the full search wins. On the other 21, only the search with forged structure succeeds. %
The control clears the base guard on $70\%$\ci{56\%}{81\%} of items and T on $52\%$\ci{39\%}{65\%}. Every win against T puts the searched text before the request, as do 22 of 35 wins on the base guard (13 put text on both sides). %
Unlike the one fixed filler of \cref{tab:ablation} ($0\%$), the control picks the best of up to 300 benign candidates per item by the guard's score. %
Each item's first candidate, drawn before any score is seen, clears the base guard on $28\%$\ci{17\%}{42\%} of items but T on only $6\%$\ci{2\%}{16\%}. %

GCG clears T on all $20$ items it ran on, raising the per-item worst case against T to $49/50$ ($98\%$\ci{89\%}{100\%}). %

\section{Related Work}
\label{sec:related}

\paragraph{Structure forgery against generators and judges.}
The ``fake completion'' and ``combined'' attacks of \citet{liu2024formalizing} forge the end of one task and the start of another; we apply that construction to guards.
In attacks descended from indirect prompt injection~\citep{greshake2023not}, forged template or conversation structure redirects agents and chat models~\citep{deng2026automating,chang2025chatinject,chen2024goal,li2025separator,ye2026prompt}, and injected content can manipulate the scores of LLM judges~\citep{shi2024optimizationbased,tamber2025illusions,li2025llms}.
Our targets instead grade the forged task, so success is a false ``safe'' verdict.

\paragraph{Attacks on guards.}
Guards fall to universal perturbations~\citep{mangaokar2024prp}, character injection~\citep{hackett2025bypassing}, pipeline attacks~\citep{mckenzie2025stack} and forged judge outputs~\citep{mccauley2025hat}.
The closest work is concurrent: \citet{li2026same} forge verdict and authority fields in a typed judge's structured input, and none of their forgeries introduces a new task to grade.

\paragraph{Structured-query and distillation defenses.}
StruQ and SecAlign~\citep{chen2024struq,chen2024secalign} combine reserved delimiters with training to ignore injected instructions.
Later work extends this line~\citep{chen2025metasecalign,chen2025defending,zhao2026ltbd} or distills the model toward its clean-context behaviour~\citep{peng2026secopd,dhouib2026raised}.
All of them defend generative instruction-followers; we apply the idea to a classifier's content slot, and our target does not need to know which segment is the injection.
Like other fine-tuning defenses~\citep{pandya2025may,nasr2025attacker}, ours falls to adaptive attacks.

\section{Limitations and Conclusion}
\label{sec:discussion}\label{sec:limitations}

\paragraph{Limitations.}
The confirmatory study covers one guard, one English corpus and short documents, and the Llama Guard~4 and role-boundary checks use 30 requests (\cref{sec:crossmodel}). Bypass counts only flagged items (Shieldstral's clean recall is $47\%$ on our clean test sample, $50\%$ on the core set), bounds end-to-end harm from above, and was not tested on retrieval, tool or multi-turn channels. %
Some results rest on small runs: the Qwen3Guard probe (40 items), the adaptive attacks (50 items; GCG on 20, not on B$+$T), and single seeds for all trained conditions but T and naive adversarial training.
Our held-out defense results cover appended forms only, and the defense only partly blunts searched benign prompts (35/50 to 26/50 items, \cref{sec:adaptive}). %
Segment-max keeps only combined harm the base guard detects, we ran no composition attack, and we claim no general prompt-injection robustness~\citep{pant2026inseparability}.

\paragraph{Conclusion.}
Guards that read their task and the content as one plain-text stream can be made to grade a forged task. Segment-max training cuts appended-attack bypass to about $1\%$, but an adaptive attacker with score access still beats it. Deployers should train the guard and never act on its verdict alone.

\section{Ethics and Disclosure}
\label{sec:ethics}

\paragraph{Disclosure.}
We reported the attack to the vendors of all five guards in October 2026, before releasing this paper, and none had responded at the time of writing.
An informal version of the attack has been public since an earlier blog post~\citep{kanwat2026blog}, so we release the paper, the benchmark and the defense without an embargo. Vendors can adopt the defense as a mitigation.

\paragraph{Harms and benefits.}
The attack is simple and transfers across guards, so we document it and offer a partial mitigation, one that nearly closes the enumerated appended channel but not adaptive attacks, instead of leaving it unmeasured.
Every harmful request we test comes from published benchmarks (\cref{sec:benchmark}). The only harmful text we wrote is eight one-line harmful decoys in the style of AdvBench goals, with no operational detail, used to show that the guard still flags a harmful last document. %

\bibliographystyle{plainnat}
\bibliography{refs}

\appendix
\crefalias{section}{appendix}%
\section{Protocol and Pre-registered Outcomes}
\label{app:protocol}

We froze the first protocol before the confirmatory run and did not edit it after seeing results, and we record every deviation from either protocol. %
These runs are exploratory and fall outside both frozen protocols: the three-decoy comparison, the off-topic-query control, the forged-verdict controls, the forged-query (attacker knowledge) variants (\cref{sec:attack}), the Qwen3Guard probe (\cref{sec:request-unit}) and the clean-copy ablation (\cref{sec:evaluation}).
The second protocol was frozen before any test-split run and has four recorded deviations. %
\begin{itemize}[leftmargin=1.5em,itemsep=1pt]
  \item D1 adds the variability design: three instantiations per item and family, two paraphrased deployment policies on six families, and three training seeds per defense condition.
  \item D2 replaces Qwen3Guard's conversation-only native forgery with the full-task forgery found on the train split, and adds the length control.
  \item D3 trims compute: 5{,}000 training items with two attacks each, three seeds for B$+$T and two for T and naive adversarial training, GCG on 20 items at 250 steps, and structural search on 50 items.
  \item D4 was recorded after a first round of test-split defense runs and before the reported ones, so it is a deviation made with test results in view. The fixed-decoy conditions of \cref{tab:fix-fixed} come from that first round. D4 draws training decoys and forged queries from one half of each pool and evaluates on the other half, adds the score-guided search and the two training-free baselines, and replaces the D3 seed plan with a smaller set to save compute: two seeds for T and one for B$+$T and the clean-copy target. %
\end{itemize}

\begin{table}[!tbp]
  \centering\footnotesize
  \caption{Pre-registered predictions and outcomes. P: confirmatory study; M: benchmark; F, A: defense. ``Refuted'': the data contradict the prediction; ``not confirmed'': the test is inconclusive. Intervals are 95\%.}
  \label{tab:prereg}
  \begin{tabular}{@{}l p{0.36\linewidth} p{0.52\linewidth}@{}}
    \toprule
    & prediction & outcome \\
    \midrule
    P1 & instruction-only $\le 10\%$ & confirmed (0\%) \\
    P2 & forged pair $\ge 70\%$ & confirmed (100\%) \\
    P3 & harmful decoy lowers bypass & not a valid test as run: the panel's ``harmful decoy'' scores $0.03$ alone, so it is the borderline decoy; bypass fell to 29--35\%; the exploratory three-decoy comparison and M3 address the prediction \\
    P4 & not recency: prepending clears markedly less than appending & confirmed (inert filler appended last 0\%; forged pair appended 100\%, prepended 15\%) \\
    P5 & syntax not load-bearing & confirmed (all variants high, no labels 0\%) \\
    P6 & wide policy-dependence & refuted: median 99\%, range 87--100\% \\
    P7 & directive does not fix & confirmed (unchanged) \\
    P8 & cross-model transfer & confirmed (30/30, 29/29) \\
    P9 & escaping fixes it & refuted: bracket-escape 88\%, label-strip 91\%, nonce-fence 93\% \\
    \midrule
    M1 & a core family $\ge 70\%$ on every guard & confirmed; best family $93.4$--$100\%$ \\
    M2 & instruction-only $\le 10\%$ on every guard & refuted on ShieldGemma ($20.2\%$); $\le 7.0\%$ on the other four \\
    M3 & dose--response, $\rho \le -0.8$ and non-increasing bins (Shieldstral, Qwen3Guard) & partly supported on Shieldstral: $\rho=-0.84$, interval $[-0.93,-0.67]$ over decoys ($[-0.85,-0.82]$ over requests), $-0.78$ on the dev split, bins not monotone; not run on Qwen3Guard \\
    M4 & control-token syntax $\ge$ angle; gap closes with special parsing off & partly supported; higher in 7 of 8 comparisons but clearly so only on Shieldstral; the parsing-off arm was not part of the benchmark run \\
    M5 & exploratory: category, source, size & released with the artifacts; size not varied within a series \\
    \midrule
    F1 & B alone leaves held-out bypass $\ge 50\%$ & confirmed; $76.8\%$\ci{75.8\%}{77.8\%} \\
    F2 & T generalises worse than B$+$T on held-out families & confirmed by the pre-registered criterion, with an effect under one point; T (mean of two seeds) $-$ B$+$T $= +0.7$ points\ci{$-0.1$}{$+1.5$}, one-sided $p=0.04$ ($n=290$); both T seeds above B$+$T on the paired items ($0.8\%$, $1.6\%$ vs $0.5\%$) \\
    F3 & B$+$T held-out bypass $\le 10\%$ & confirmed; $0.9\%$\ci{0.7\%}{1.1\%} \\
    F4 & B$+$T AUROC within $0.01$, recall within 2 points (on the interval of the difference) & not confirmed; AUROC $-0.010$\ci{$-0.041$}{$+0.021$}, recall $-0.1$\ci{$-5.0$}{$+4.8$}: point estimates inside the margins, intervals wider \\
    F5 & B$+$T over-refusal within 2 points (on the interval of the difference) & not confirmed; XSTest-safe $-2.7$\ci{$-11.1$}{$+5.4$}, OR-Bench-hard $+0.25$\ci{$-6.6$}{$+7.1$}, structured benign $0.0$\ci{$-3.0$}{$+3.0$}: no point estimate rises by more than $0.25$, intervals wider than the margin \\
    F6 & naive adversarial training costs $\ge 3$ points structured-benign over-refusal or generalises worse & refuted with the fixed decoy ($0.0\%$ on a set overlapping its hard negatives, $0.9$--$1.1\%$ held-out); its costs are clean FPR and the trailing-pair blind spot \\
    \midrule
    A1 & GCG against B$+$T $\le 30\%$ & not tested on B$+$T; on T GCG clears 20/20 \\
    A2 & structural search against B$+$T $\le 20\%$ & confirmed; 4/50 ($8\%$\ci{3\%}{19\%}); the score-guided search, added in D4 without a prediction, clears 34/50 \\
    A3 & LLM attacker & not run \\
    \bottomrule
  \end{tabular}
\end{table}

\section{Benchmark Grammar}
\label{app:grammar}

The attack grammar has the axes below. The full grammar has 654 attack forms and 5 controls, one of which is the no-attack condition. Its core panel has 63 forms: 58 attack forms in 52 families (five guard-native families have two or three forms each) and the same 5 controls. %
A family is held out if its syntax is JSON, YAML, markdown bold or control token, if its language is neither English nor Spanish, or if it has at least three forged pairs. Spanish-label families are therefore held out only through the syntax or pair-count rule. The defense is trained on the remaining families. %
\begin{itemize}[leftmargin=1.5em,itemsep=1pt]
  \item label syntax: guard-native tags, angle brackets, square brackets, markdown, plain colon, abbreviations, JSON, XML, YAML, control tokens;
  \item number of forged pairs: 1--5; position: append, prepend, middle;
  \item decoy harm: benign, borderline, harmful; forged query: exact, paraphrase, generic, out-of-domain; language; encoding: plain text or real control tokens.
\end{itemize}
Per-family, per-category and per-source bypass and threshold sensitivity are released with the artifacts.

\section{Additional Attack Analyses}
\label{app:extra}

\paragraph{Activation patching.}
Activation patching on Shieldstral does not locate the mechanism, and we treat it as uninformative about rebinding.
We run the attacked input and the same input with request and decoy swapped ($n=100$), using for each request a decoy of the closest token length and matching positions by role, not by index, which keeps any remaining length mismatch from misaligning them.
Patching the residual stream at the final position flips the verdict on at least $99\%$ of inputs at every layer from 20 onward (of 26, numbered from 0), which shows only that the verdict is read out there. %
Patching the forged labels, alone or with the forged query between them, at any layer moves the yes/no logit margin by at most $0.05$ of the gap between the two inputs and flips at most $3\%$ of verdicts. %
These positions hold the same text in both runs; we did not patch the request or decoy positions, so the experiment does not test whether the guard grades the last document. %

\paragraph{Benchmark variations.}
Real control-token syntax changes bypass relative to angle-bracket labels by $-1.5$ to $+10.4$ points, which only partly supports our prediction that it would raise bypass. %
Across the three deployment policies, bypass moves by at most $6.8$ points on Shieldstral, Llama Guard~3 and Qwen3Guard, and by up to $13.7$ points on ShieldGemma. %

\paragraph{Multiple verdicts from gpt-oss-safeguard.}
On stacked forged pairs gpt-oss-safeguard sometimes returns one verdict per content block.
We count an item as flagged if any verdict is 1, the reading least favourable to the attack, chosen after the multi-verdict replies were seen. %
Reading only the last verdict, as a pipeline acting on the trailing label would, raises two stacked-pair families from $2.9$--$3.5\%$ to $87.3$--$90.6\%$ and moves every other family by under one point. %

\section{Defense Implementation and Training}
\label{app:defense-impl}

\paragraph{Bounded format.}
The template puts a space before the document, which the tokenizer would otherwise merge into the content's first token. We move that space into the trusted prefix, so plain documents tokenize exactly as they do in the stock format (verified on 1{,}600 documents: AdvBench rows 200--299 and the clean test sample). %

\paragraph{Training.}
We train a LoRA (rank 16, $\alpha=32$) on the language model's attention and MLP projections with AdamW at learning rate $10^{-4}$, batch 8, one epoch and maximum length 384. LoRA does not touch the embeddings. In B$+$T a trainable two-row offset, initialised at zero, is added to the embeddings of the two boundary tokens (ids 36 and 37), and every other embedding stays frozen. %
The data are 5{,}000 train-split prompt items, each clean and with two forms sampled uniformly from the training branch, plus 1{,}250 structured-benign items, for 16{,}250 examples. Of the 10{,}000 attacked examples, 9{,}480 append the forged structure, 317 prepend it or use the middle position, and 203 are controls with no forged structure. %
 One epoch is 2{,}032 steps, the last batch partial. Each attacked example's decoy and forged query are a seeded draw from the training halves of the pools: 20 of the 40 benign decoys, 3 of 6 borderline and 4 of 8 harmful decoys, and 4 of the 8 forged queries at each relatedness level. %
Naive adversarial training and the clean-copy ablation use the same examples with different targets.

\paragraph{Compute and adaptive budgets.}
Training and the test-split defense, adaptive and mechanism runs used one NVIDIA RTX 4090. A training run takes about 17 minutes (1{,}028\,s for T and 1{,}037\,s for B$+$T at the last logged step, 2{,}030 of 2{,}032). The benchmark ran on an Apple-silicon laptop GPU. %
The score-guided search has 300 guard calls per item on 50 items, starting from 20 random candidates and then mutating batches of 10 children with restart probability 0.2. The structural search scores all 684 attacks on each of 50 items (batch 128). GCG optimises a 20-token suffix inside the forged document for 250 steps (batch 32 candidates per step) on 20 items, and stops on an item at its first success. %

\paragraph{Fixed-decoy conditions.}
\Cref{tab:fix-fixed} gives the conditions trained and tested with one fixed decoy and query per family: naive adversarial training and the generality runs of T on Llama Guard~3 1B and Qwen3Guard 0.6B (\cref{sec:evaluation}).
Their training drew hard negatives from all splits, not only the train split, and these include 380 of the 500 structured-benign test items, so the structured-benign column of the trained rows is not a held-out measure. %

\begin{table}[ht]
  \centering
  \footnotesize
  \caption{Fixed-decoy conditions on the test split, $\tau=0.5$, in \%. Columns as in \cref{tab:fix-main,tab:fix-utility}; naive adversarial training has two seeds (range), the other rows seed~0.}
  \label{tab:fix-fixed}
  \setlength{\tabcolsep}{2pt}
  \begin{tabular}{l cc ccc ccc cc}
    \toprule
    & \multicolumn{2}{c}{bypass $\downarrow$} & \multicolumn{3}{c}{detection} & \multicolumn{3}{c}{over-refusal $\downarrow$} & \multicolumn{2}{c}{benign $+$ forged pair} \\
    \cmidrule(lr){2-3}\cmidrule(lr){4-6}\cmidrule(lr){7-9}\cmidrule(lr){10-11}
    condition & train & held-out & AUROC & recall & FPR & OR-B. & XST. & struct. & benign & harmful \\
    \midrule
    naive adv.\ training (2 seeds) & 0.9--1.0 & 0.9--1.1 & 0.921--0.925 & 83.3--84.6 & 12.6--14.9 & 0.3--0.8 & 8.0--10.7 & 0.0 & 7.8--8.2 & 8.0--8.6 \\
    Llama Guard 3 1B, base & 84.0 & 82.0 & 0.758 & 58.7 & 16.9 & 40.5 & 1.3 & 23.4 & 2.4 & 97.3 \\
    Llama Guard 3 1B, T    & 2.3 & 2.5 & 0.771 & 58.4 & 15.0 & 38.3 & 1.3 & 23.0 & 22.5 & 100 \\
    Qwen3Guard 0.6B, base  & 28.3 & 9.5 & 0.917 & 88.5 & 23.6 & 76.8 & 21.3 & 30.4 & 10.5 & 100 \\
    Qwen3Guard 0.6B, T     & 0.0 & 0.0 & 0.924 & 89.0 & 23.3 & 77.3 & 20.0 & 29.4 & 25.9 & 100 \\
    \bottomrule
  \end{tabular}
  \par\smallskip\raggedright OR-B.: OR-Bench-hard; XST.: XSTest-safe; struct.: structured benign; benign / harmful: decoy of the forged pair on benign content.
\end{table}

\paragraph{Content tokens.}
Disabling special-token parsing leaves atomic any added token that the tokenizer does not mark special~\citep{zhan2026samebytes,yang2026nameless}, so we checked which added-vocabulary ids the content encoder can still emit.
For each guard tokenizer we encoded every added token's string on its own, embedded in plain text and in 500 random concatenations, with special-token parsing disabled, and recorded any added or special id in the output. %
No added or special id is reachable from content on Shieldstral, the guard the defense is trained on (1000 added tokens, all special, with the boundary tokens at ids 36 and 37), or on Llama Guard~3 1B (256, all special).
On Qwen3Guard-0.6B, 12 of 26 added tokens are not marked special and stay atomic, among them the tool-call tags \texttt{<tool\_call>} and \texttt{</tool\_call>} and the fill-in-the-middle markers. On ShieldGemma 2B, 242 of 249 are not special (the \texttt{<unused$k$>} and other reserved markers) and also stay atomic.
Content can produce these ids, so a boundary-token design on these tokenizers must not use them as boundary tokens.

\section{Reproducibility}
\label{app:repro}

The runs used these Hugging Face revisions: Shieldstral-1.0-3B \texttt{003ec7e}, Llama Guard~3 1B \texttt{acf7aaf}, Qwen3Guard-Gen-0.6B \texttt{fada3b2} and ShieldGemma 2B \texttt{d1dffc9}. Local guards are scored from a single forward pass with no sampling. The score is the probability of the unsafe label renormalised over the guard's label tokens, computed in fp32 from bf16 hidden states. We reached the API-served guards, gpt-oss-safeguard-20B and Llama Guard~4 12B, through OpenRouter at temperature 0 without pinning a provider. Every benchmark call to gpt-oss-safeguard was served by Groq, as recorded in the response metadata; the cross-model runs did not record the provider. %
Dataset sources and licenses: AdvBench, HarmBench, Aegis~2.0, BeaverTails, the OpenAI moderation evaluation and ToxicChat, with XSTest and OR-Bench as over-refusal sources. We use the non-commercial sets (ToxicChat, BeaverTails) for research only. %

\section{Open Science}
\label{app:open-science}

We release the \guardrebind{} benchmark and harness, the attack grammar, the defense code, the training commands and data recipe, and the result files behind every number in this paper, at \url{https://github.com/ukanwat/guardrebind}. We also release a T adapter for Shieldstral at \url{https://huggingface.co/ukanwat/shieldstral-1.0-3b-rebinding-fix}. Its weights come from a retraining with the same recipe and seed as T seed~0, on newer library versions; on the held-out-decoy measure it is cleared on $0.95\%$ ($67/7{,}032$) of attacked inputs, against $0.81\%$ for the run reported here, and its AUROC is $0.799$. %
The released results contain item ids and scores, plus text only from the public benchmarks we used (such as the AdvBench requests of the confirmatory run and the benign development-split prompts that won as search decoys) and model outputs that quote it. We release no new harmful content.
The pre-registered protocol and its deviations are included so that predictions can be checked against outcomes.

\end{document}